\documentclass[preprint,aip,apl]{revtex4-1}
\usepackage[utf8]{inputenc}
\DeclareUnicodeCharacter{FB01}{fi}

\usepackage{amsmath}    
\usepackage{graphicx}   
\usepackage{verbatim}   
\usepackage{color}      
\usepackage{epstopdf}  
\usepackage{upgreek} 
\usepackage[hidelinks]{hyperref}   
\usepackage{textcomp} 
\usepackage{siunitx}

\newcommand{\Neel}{Néel}
\newcommand{\MuMax}{MuMax$^{3}$ }
\newcommand{\figref}[1]{Fig.~\ref{#1}}

\begin{document}

\title{Tuning magnetic chirality by dipolar interactions}
\author{Juriaan Lucassen}
\email{j.lucassen@tue.nl}
\affiliation{Department of Applied Physics, Eindhoven University of Technology, 5600 MB Eindhoven, the Netherlands}
\author{Mariëlle J. Meijer}
\affiliation{Department of Applied Physics, Eindhoven University of Technology, 5600 MB Eindhoven, the Netherlands}
\author{Fabian Kloodt-Twesten}
\author{Robert Fr\"{o}mter}
\affiliation{Universität Hamburg, Center for Hybrid Nanostructures, Luruper Chaussee 149, 22761 Hamburg, Germany}
\author{Oleg Kurnosikov}
\affiliation{Department of Applied Physics, Eindhoven University of Technology, 5600 MB Eindhoven, the Netherlands}
\author{Rembert A. Duine}
\affiliation{Department of Applied Physics, Eindhoven University of Technology, 5600 MB Eindhoven, the Netherlands}
\affiliation{Institute for Theoretical Physics, Utrecht University, Leuvenlaan 4, 3584 CE Utrecht, the Netherlands}
\author{Henk J.M. Swagten}
\affiliation{Department of Applied Physics, Eindhoven University of Technology, 5600 MB Eindhoven, the Netherlands}
\author{Bert Koopmans}
\affiliation{Department of Applied Physics, Eindhoven University of Technology, 5600 MB Eindhoven, the Netherlands}
\author{Reinoud Lavrijsen}
\affiliation{Department of Applied Physics, Eindhoven University of Technology, 5600 MB Eindhoven, the Netherlands}

\date{\today}
\begin{abstract}
Chiral magnetism has gained enormous interest in recent years because of the anticipated wealth of applications in nanoelectronics. The demonstrated stabilization of chiral magnetic domain walls and skyrmions has been attributed to the actively investigated Dzyaloshinskii–Moriya interaction. Recently, however, predictions were made that suggest dipolar interactions can also stabilize chiral domain walls and skyrmions, but direct experimental evidence has been lacking. Here we show that dipolar interactions can indeed stabilize chiral domain walls by directly imaging the magnetic domain walls using scanning electron microscopy with polarization analysis. We further show that the competition between the Dzyaloshinskii–Moriya and dipolar interactions can reverse the domain-wall chirality. Finally, we suggest that this competition can be tailored by a Ruderman–Kittel–Kasuya–Yosida interaction. Our work therefore reveals that dipolar interactions play a key role in the stabilization of chiral spin textures. This insight will open up new routes towards balancing interactions for the stabilization of chiral magnetism.
\end{abstract}

\maketitle
The role of chirality is becoming more important for new applications in spintronics, especially in ultrathin magnetic films.~\cite{Bode2007,Ryu2013,Emori2013,PhysRevLett.87.037203,doi:10.1063/1.5048972,Fert2017} In magnetic racetrack applications for example, the chirality directly determines how magnetic domain walls and skyrmions interact with the spin-orbit torques.~\cite{Ryu2013,Emori2013,Fert2013,Legrandeaat0415,PhysRevB.98.104402,PhysRevApplied.10.064042} It is therefore important to investigate the key contributing factors to this chirality. The underlying interaction that is believed to stabilize the chirality is the Dzyaloshinskii–Moriya interaction (DMI). As shown by a wealth of theoretical and experimental reports this interaction requires the breaking of inversion symmetry and originates from the interface between a heavy metal and a ferromagnet for the thin film systems investigated in this paper.~\cite{PhysRevLett.87.037203,PhysRevB.78.140403} The DMI also helps to stabilize skyrmions because it favours non-collinear spin configurations,~\cite{Nagaosa2013} which are envisaged to be used in areas ranging from magnetic racetrack memory and logic applications, to radio frequency devices and neuromorphic computing.~\cite{doi:10.1063/1.5048972,Fert2017} 

Very recently, however, it was realized that DMI is not the only interaction that can stabilize a specific chirality.~\cite{PhysRevApplied.10.064042,Legrandeaat0415,PhysRevB.98.104402,Dovzhenko2018,2019arXiv190103652F,Hrabec2017} Actually, already $40$~years ago it was shown that the presence of dipolar fields leads to the formation of chiral \Neel{} caps.~\cite{HubertDomains,malozemoff1979magnetic} Here, the stray fields originating from magnetic domains align the spins inside the domain walls at the top of the film to form clockwise (CW) \Neel{} walls and and at the bottom of the film to form counterclockwise (CCW) \Neel{} walls, providing an optimized flux closure state. Dipolar interactions can often be ignored for the thin-film systems used for domain-wall studies. Because of the increase in magnetic volume and reduced coupling across the non-magnetic spacer layers this is no longer the case for the multilayer repeat systems often used to stabilize room-temperature magnetic skyrmions.~\cite{Luchaire_skyrmion,Woo2016,Hrabec2017,HubertDomains,malozemoff1979magnetic} 

Both theoretical~\cite{PhysRevApplied.10.064042,Legrandeaat0415,PhysRevB.98.104402} and experimental work~\cite{Legrandeaat0415,Dovzhenko2018,2019arXiv190103652F} suggests that in these multilayer repeat systems the DMI is in direct competition with the dipolar fields. Without DMI, the dipolar interactions introduce \Neel{}~caps. Including DMI, however, leads to a larger fraction of the layers being occupied by the \Neel{} cap of the chirality favoured by the DMI. The other cap will be reduced in size and occupy fewer layers. This happens until the DMI is so large that it is no longer energetically favourable to accommodate a \Neel{} cap not favoured by the DMI. 

The energetics and dynamics of both skyrmions and domain walls are affected by this competition because it determines the net chirality of the magnetic textures, which in turn influences the interaction with, for example, spin-orbit torques.~\cite{Legrandeaat0415,PhysRevB.98.104402,PhysRevApplied.10.064042} A fundamental understanding of this competition is therefore needed to properly tailor the interactions such that we get the desired domain-wall and skyrmion profiles. Theoretically, the situation has been made increasingly clear over the past two years~\cite{PhysRevB.95.024415,Legrandeaat0415,PhysRevB.98.104402,PhysRevApplied.10.064042,Dovzhenko2018}, but unambiguous experimental verification of the predicted domain-wall behaviour has proven to be extremely challenging because it is very difficult to image magnetic domain walls directly. Observations reported in Refs.~\onlinecite{Dovzhenko2018}~and~\onlinecite{2019arXiv190103652F} suggest the presence of \Neel{} caps based on measurements of the magnetic stray fields. Another study directly measures the domain-wall chirality across a range of systems and finds one stack out of many where the domain-wall chirality appears to be determined by the dipolar interactions.~\cite{Legrandeaat0415} However, direct systematic evidence highlighting the competition between both interactions and the ability of both to stabilize chiral \Neel{} walls is lacking. 

In this article we therefore explicitly address the competition between the effective DMI, characterized by the constant $D$, and dipolar interactions for the formation of chiral magnetic domain walls. We do this by directly imaging the magnetic domain-wall texture in the top CoB layer of a Pt/CoB/Ir multilayer repeat system using a scanning electron microscope with polarization analysis (SEMPA).~\cite{Oepen2005,*UNGURIS2001167} By varying the thickness of the top CoB layer we tune the strength of the effective DMI of this top layer and leave the dipolar fields from the bottom repeats unaffected. We find that going from low ($<1.1$~\si{nm}) to high ($>1.1$~\si{nm}) CoB thicknesses the domain-wall chirality reverses from CCW \Neel{} to CW \Neel{} because we tune the effective DMI, which originates from the Pt/CoB interface, with respect to the dipolar interactions. We elucidate the physical interactions at play using a simple analytical model and we explicitly demonstrate the validity of our interpretation using micromagnetic simulations. Additionally, we argue that a small indirect (asymmetric~\cite{2018arXiv180901080H,*2018arXiv181001801F}) exchange interaction, known as the Ruderman–Kittel–Kasuya–Yosida (RKKY) interaction, that couples magnetic layers across a non-magnetic spacer can tailor the competition between the dipolar interaction and the DMI. Our results pave the way towards engineering of the magnetic chirality in thin film magnetic systems.

\begin{figure*}
\centering
\includegraphics[width=1\textwidth]{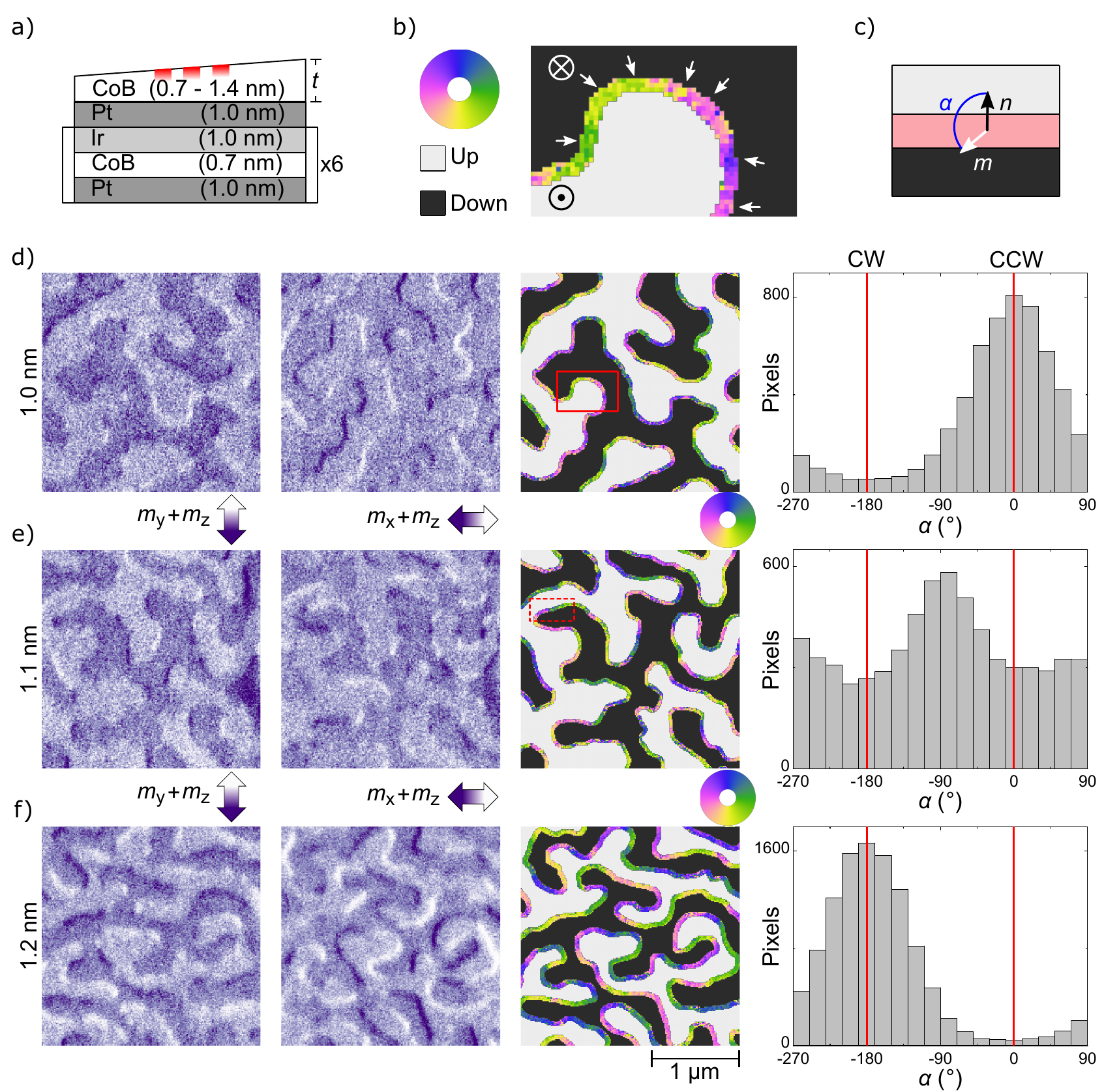}
\caption{\label{fig:figure1} \fontsize{0.34cm}{0.4cm}\selectfont a) Schematic cross section of the investigated sample. Indicated in red are the investigated areas on the wedge sample corresponding to a thickness $t$ of the top CoB layer of $1.0$, $1.1$ and $1.2$~\si{nm}. b) Zoom in composite image of the area indicated in red in d). The OOP domains are indicated in black and white and the in-plane magnetic domain wall is shown in color, with the color corresponding to the direction of the magnetization in the wall as given by the color wheel. We also denote the magnetization directions with arrows. c) Definition of the domain-wall angle $\alpha$ between the domain-wall normal $n$ and the magnetization inside the domain wall $m$. d-f) SEMPA images and analysis for different $t$ of $1.0$, $1.1$ and $1.2$~\si{nm}. From left to right: 1. $m_\mathrm{y}$ + OOP contrast ($m_\mathrm{z}$). 2. $m_\mathrm{x}$ + OOP contrast ($m_\mathrm{z}$). The arrows indicate the direction of the magnetization contrast. 3. composite image constructed from 1. and 2 [see b)]. 4. Histogram of the domain-wall angle $\alpha$ [see c)] for all pixels in the domain wall where $\alpha=0,180^{\circ}$ indicates a counterclockwise (CCW) and clockwise (CW) \Neel{} wall respectively. In e) the dashed box indicates a Bloch line in the domain wall. The scale bar for all SEMPA images is shown at the bottom.}
\end{figure*}

The investigated system is as follows: //Ta(4)/Pt(2)/\allowbreak[Pt(1)/Co$_{80}$B$_{20}$(0.7)/Ir(1)]x6\allowbreak/Pt(1)\allowbreak/Co$_{80}$B$_{20}$($t$) (thicknesses in parentheses in \si{nm}) which was DC sputter deposited using an Ar pressure of~$2\times 10^{-3}$~\si{mbar} on a Si substrate with a native oxide in a system with a base pressure of $2\times 10^{-9}$~\si{mbar}. We use Pt and Ir because of their expected high and additive DMI which favours CCW \Neel{} walls ($D>0$),~\cite{PhysRevLett.120.157204,*PhysRevLett.118.147201,*Han2016,*Finizio2019,Luchaire_skyrmion} and we use CoB because it is easy to obtain an as-deposited multidomain state necessary for domain wall imaging. An overview of the stack is displayed in~\figref{fig:figure1}a, where we wedged the top layer such that $t$ varied continuously between $0.7$ and $1.4$~\si{nm} to directly access a regime of low and high effective $D$ in a single layer due to the interfacial nature of the DMI. After deposition, the sample was transported \textit{in-situ} to our SEMPA chamber. SEMPA combines the ability to map both in-plane (IP) magnetization components ($m_\mathrm{x}$ and $m_\mathrm{y}$) simultaneously with an extreme surface sensitivity.~\cite{Oepen2005,*UNGURIS2001167,doi:10.1063/1.4998535} We tilted the sample slightly~\footnote{The tilt is around $5^\circ$, varying slightly depending on the CoB thickness} to create out-of-plane (OOP, $m_\mathrm{z}$) contrast in both $m_\mathrm{y}$ and $m_\mathrm{x}$ such that we image both the domain walls and the OOP domains simultaneously, which allows us to extract the chirality of the top CoB layer.~\cite{doi:10.1063/1.4998535,PhysRevLett.100.207202,PhysRevB.96.060410}  For $t \geq 1.4$~nm, the top CoB layer turns IP and starts to align along the stray fields of the underlying layer.~\cite{doi:10.1063/1.4998535}  For the micromagnetic simulations, we used \MuMax (see supplementary note II)~\cite{Vansteenkiste2014} and the details of the analytical calculations can be found in supplementary note III.

A SEMPA measurement on a CoB thickness of $t=1.0$~\si{nm} is shown in~\figref{fig:figure1}d. The first two images show the $m_\mathrm{y}$ and $m_\mathrm{x}$ magnetization components imaged simultaneously on the same area, where both images also contain OOP ($m_\mathrm{z}$) contrast.~\footnote{The OOP component present in the $m_\mathrm{x}$ image is not properly defined and depends sensitively on the sample mounting conditions} There are two main features in the $m_\mathrm{y}$ image. First, a worm-like domain pattern given by the OOP domains. Second, a dark lining on top of the light domains, and a bright lining on the bottom of the light domains, which correspond to the domain walls. This lining is not present on the left/right of the domains, which suggests these are \Neel{} walls. In the $m_\mathrm{x}$ measurement the linings are on the left and right side of the OOP domains. We can combine both the OOP and IP information from both images to form a composite image~\citep{doi:10.1063/1.4998535,Chen2013,Note5}~\footnotetext[5]{This was done by hand} which is the third image in~\figref{fig:figure1}d and a zoom of this image is shown in \figref{fig:figure1}b. 

In this zoomed image, we indicate the OOP domains in white and black, inferred from the contrast in the $m_\mathrm{y}$ image. Superimposed, we show the domain walls where the direction of the magnetization inside the domain walls is indicated in color following the colorwheel. As signalled by the arrows, the magnetization inside the wall points from down to up domains, indicating a CCW \Neel{} wall.~\cite{PhysRevB.78.140403,Chen2013,doi:10.1063/1.4998535} A more thorough look at the composite image of~\figref{fig:figure1}d suggests this is true for most domain walls at this thickness. To quantify this further, we construct a histogram of the domain-wall angle $\alpha$ between the domain-wall normal $n$ and the domain-wall magnetization $m$ (see~\figref{fig:figure1}c for a sketch) which is shown on the far right in~\figref{fig:figure1}d. This histogram is sharply peaked around $\alpha=0^\circ$ which indicates that CCW \Neel{} walls dominate at this thickness. The full width at half maximum (FWHM) of the histogram could be used to determine the spread in domain-wall angles. However, we estimate that the dominating contribution to the FWHM of the histogram is the result of Poisson noise in the electron counting~\cite{doi:10.1063/1.3534832} and errors in the determination of the domain-wall normal (see supplementary III).~\cite{doi:10.1063/1.4998535} We therefore give no quantitative estimation of the spread in domain-wall angles, but we know it to be much smaller than the histogram width.

In~\figref{fig:figure1}e and f we show the same measurement and analysis for $t=1.1$ and $1.2$~\si{nm}, respectively. Measurements for different thicknesses on the same sample and similar measurements on a nominally identical sample can be found in supplementary note I. For $t=1.2$~\si{nm} (\figref{fig:figure1}f) the domain-wall chirality has reversed and the corresponding histogram is now peaked around a CW \Neel{} wall orientation. For $t=1.1$~\si{nm} (\figref{fig:figure1}e) there is a transition region with different types of domain walls. From the histogram we conclude that there is slight preference for Bloch walls based on the minimal peaks shown at $\alpha=\pm 90^\circ$. Additionally, we also find Bloch lines inside the walls, where the magnetization direction in the wall suddenly rotates by $180^\circ$ (see dashed rectangle).~\footnote{We also note a decreasing domain size with increasing CoB thickness that we speculate to be the result of an as-deposited state that is not the ground state of the system.} This transition from a CW \Neel{} wall at $1.2$~\si{nm} to a CCW \Neel{} wall at $1.0$~\si{nm} is the result of dipolar interactions that are in direct competition with the DMI. Below $1.1$~\si{nm} the DMI dominates whereas dipolar interactions dominate above $1.1$~\si{nm}. 

\begin{figure}
\centering
\includegraphics{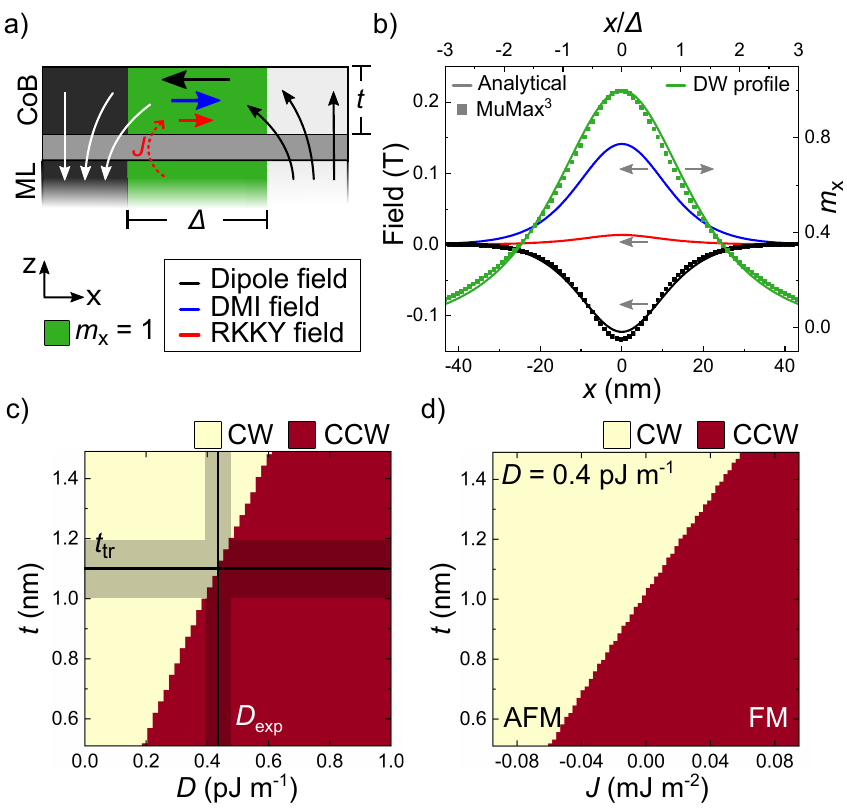}
\caption{\label{fig:figure2}a) Basics of the model. A CoB layer on top of a multilayer (ML) stack with CCW \Neel{} walls indicated in green with the $3$ effective IP magnetic fields acting on the domain wall in this layer indicated by the arrows. b) Magnitude of the effective fields (black, blue, red; left axis) in the $x$-direction, including the domain-wall profile (green; right axis), as a function of position. Plotted are the analytical results as well as the micromagnetic results for $t=1.2$~\si{nm}, $J=0.02$~\si{mJ.m^{-2}}, and $D=0.6$~\si{pJ.m^{-1}}. c,d) Phase diagrams according to the analytical model for the domain-wall chirality in the top CoB layer as a function of: c) $D$ and $t$ with $J=0$~\si{mJ.m^{-2}} and: d) $J$ and $t$ for $D=0.4$~\si{pJ.m^{-1}}. In c), the black area indicates the experimental thickness transition  region $t_\mathrm{tr}$ and DMI $D_\mathrm{exp}$.}
\end{figure}

To substantiate the observations, in what follows we will use a simple analytical model to explain our results. The fundamentals of this model, adapted from Ref.~\onlinecite{Legrandeaat0415}, are given in~\figref{fig:figure2}a, where we show a CoB layer on top of a multilayer stack and the effective IP magnetic fields acting on the domain wall in this top CoB layer. In the original model of Ref.~\onlinecite{Legrandeaat0415} there are two in-plane magnetic fields that determine the domain-wall chirality in the top CoB layer. First, the dipolar field from the stack underneath. Second, an effective field from the DMI of the top CoB layer itself. We add to this model a third term; an RKKY interaction as Ir is a well-known RKKY mediating layer~\cite{PhysRevLett.67.3598} whose coupling can persist through thin layers of Pt.~\cite{doi:10.1063/1.3682103} This term is often overlooked but can be of significant importance as we demonstrate later in this paper. We assume that the underlying layers of thickness $0.7$~\si{nm} contain CCW \Neel{} walls induced by the DMI ($D>0$), because the experiments consistently shows CCW \Neel{} walls below $1.1$~\si{nm} (see supplementary note I). We use analytical expressions derived in supplementary note III, based on derivations presented elsewhere, to calculate the dipolar and DMI fields.~\cite{Legrandeaat0415,PhysRevB.95.174423} The details on the calculations for the added RKKY field are also presented in supplementary note III.
 
These calculated magnetic fields are shown in~\figref{fig:figure2}b for ferromagnetic ($J>0$) RKKY interaction. Apart from the strength of $D$ and $J$, which we vary, all other input parameters are based on the experimental parameters of the investigated system. The DMI and RKKY field point in the $+x$ direction ($>0$), indicating a preference for CCW \Neel{} walls. However, the dipolar field points from the up domain towards the down domain and is thus directed in the opposite ($-x$) direction favouring CW \Neel{} walls. The sum of these magnetic fields integrated across the domain-wall profile (also indicated in~\figref{fig:figure2}b) determines the eventual domain-wall chirality.~\cite{Legrandeaat0415} If the total integrated field points in the $+x$ direction, the resulting domain-wall profile will be of CCW \Neel{} origin, and if the field points in the $-x$ direction, the resulting domain-wall profile will be CW \Neel{}. In this example, the DMI and RKKY are dominant and the resulting domain-wall profile of the top CoB layer will be CCW. 

Comparing the values of the integrated magnetic fields as a function of $t$ and $D$ for a situation without RKKY coupling yields the phase plot depicted in~\figref{fig:figure2}c, where we plot the resulting domain-wall chirality of the top layer CoB as a function of both parameters. With increasing $t$ the effective DMI field reduces as $1/t$ due to its interfacial nature, until it is so small that the dipolar interactions become dominant and the top domain wall is of the CW \Neel{} type. However, if we include the RKKY interaction the situation is modified as we demonstrate in~\figref{fig:figure2}d, where the domain-wall chirality is shown as a function of both $t$ and $J$ for $D=0.4$~\si{pJ.m^{-1}}. Apparently, the transition thickness from CW to CCW can be shifted as a function of the RKKY interaction, where it shifts to thicker layers for ferromagnetic ($J>0$) coupling and to thinner layers for antiferromagnetic ($J<0$) coupling. As this effect occurs for reasonably small values of $J$~\cite{PhysRevLett.67.3598,doi:10.1063/1.3682103} we conclude that an independent quantification of both $J$ and $D$ is not possible when both interactions are present. 

We demonstrate in supplementary note V that for our samples the RKKY coupling is $|J|<0.001$~\si{mJ.m^{-2}}. Following the phase diagram, we then find that its influence is negligible and that we can use~\figref{fig:figure2}c to determine the DMI. As found above, the transition thickness $t_\mathrm{tr}$ between CW and CCW \Neel{} walls is between $1.0$~\si{nm} and $1.2$~\si{nm} experimentally. Based on these elementary model calculations we therefore conclude that the experimental DMI $D_\mathrm{exp}$ is $+0.44 \pm 0.05$~\si{pJ.m^{-1}} (indicated by the black lines in the figure) for the top Pt/CoB interface.~\footnote{Value is the mean and the error bar is half the width of the rectangle indicated in~\figref{fig:figure2}c} This value is slightly lower than reported for Pt/Co, Pt/CoFeB and Pt/Co$_\mathrm{68}$B$_{\mathrm{32}}$ interfaces, which is around $+1$~\si{pJ.m^{-1}}.~\cite{PhysRevLett.120.157204,*PhysRevLett.118.147201,*Han2016,*Finizio2019}

\begin{figure}
\centering
\includegraphics{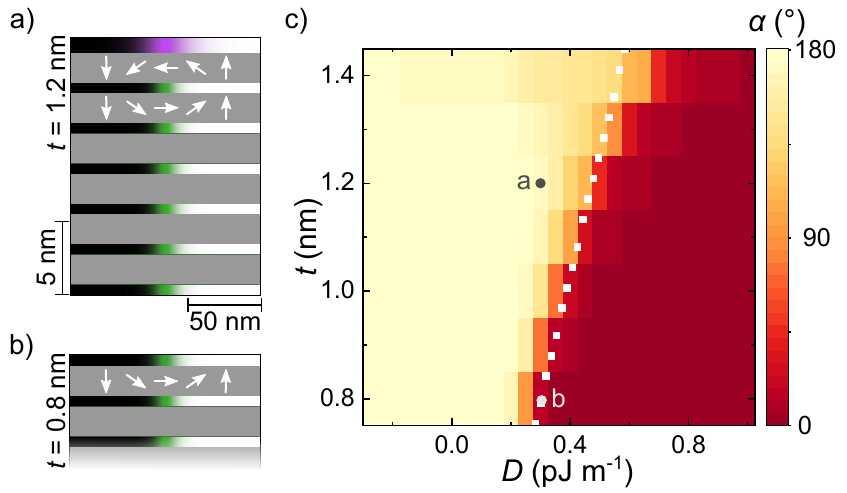}
\caption{\label{fig:figure3}a, b) Result of a micromagnetic simulation with $D=0.3$~\si{pJ.m^{-1}} for~$t=1.2$~\si{nm} (a) and $t=0.8$~\si{nm} (b). The arrows indicate the magnetization direction of the layer above. For the colors, see~\figref{fig:figure1}b. c) Phase diagram for the domain-wall angle $\alpha$~(\figref{fig:figure1}c) in the top CoB layer from micromagnetics as a function of $D$ and $t$ with $J=0$~\si{mJ.m^{-2}}. Also indicated is the transition line from CW to CCW from~\figref{fig:figure2}c and the location of the simulations shown in a-b).}
\end{figure}

Although our simple interpretation using this elementary model describes our observations quite well, two issues need to be carefully addressed. First, for these calculations we assumed the domain-wall width to be constant as a function of $t$ and matched it to a micromagnetic simulation for the parameters of~\figref{fig:figure2}b. Yet, we know this width to be thickness dependent through $K_\mathrm{eff}$ and complex interactions with the dipolar fields. Second, the model does not suggest the presence of Bloch domain walls which is what we observe experimentally at $t=1.1$~\si{nm}. To tackle these issues we derived a more complete picture by performing micromagnetic simulations where we simulate the complete system and look at its influence on the chirality of the domain walls in the top CoB layer. Using this approach, the variation in the domain-wall width and possible presence of Bloch wall is taken into account. 

In~\figref{fig:figure3}a and b we show cross sections of the magnetic texture determined using micromagnetic simulations, where the domain walls in the bottom repeats are aligned along the CCW direction favoured by the DMI (which was assumed implicitly for the analytical calculations). However, in a) due to the increased thickness ($1.2$~\si{nm}) of the top layer the effective DMI of that layer is not strong enough to overcome the dipolar interactions, resulting in a CW \Neel{} wall for the top CoB layer. In b) the top layer is thinner ($0.8$~\si{nm}) and now the effective DMI is dominant. Running these simulations for different $D$ and $t$ and extracting the resulting top domain-wall angle $\alpha$ produces~\figref{fig:figure3}c. We find behavior akin to~\figref{fig:figure2}c, where the domain-wall chirality reverses between CCW ($\alpha=0^\circ$) and CW ($\alpha=180^\circ$) as a function of $D$ and $t$. Additionally, there is now a transition region where we find Bloch domain walls ($\alpha=90^\circ$), in agreement with our experimental findings at $1.1$~nm (see~\figref{fig:figure1}). As indicated by the white dots in~\figref{fig:figure3}c, where we show the transition from CW to CCW from~\figref{fig:figure2}c, the behaviour matches quantitatively with the analytical calculations of~\figref{fig:figure2}c. As a function of $t$ we find slight deviations, which are attributed to a variation in domain-wall width as we demonstrate in supplementary note III. In supplementary note VII we further demonstrate that the experimentally determined domain-wall width also matches the domain-wall width extracted from micromagnetic simulations and that it depends on the CoB thickness $t$ as expected. Similar simulations as a function of $J$ and $t$ can be found in supplementary note VI, where we also find agreement between the analytical calculations and the micromagnetic simulations. 

Even though the agreement between the analytical calculations and simulations is excellent, we signal two issues that may need further attention. First, for small values of $D$ ($<0.15$~\si{pJ.m^{-1}} at $t=1.0$~\si{nm}), the simulations show the formation of a \Neel{} cap with CW \Neel{} walls in the underlying layers.~\cite{PhysRevApplied.10.064042,Legrandeaat0415,PhysRevB.98.104402,Dovzhenko2018,2019arXiv190103652F} This invalidates a critical assumption of the analytical model; namely that the underlying repeats are always assumed to be CCW. Therefore, in this situation we need a more complicated implementation of the RKKY interaction and dipolar fields in the analytical model. Second, experimentally we find that at $1.4$~\si{nm} the top CoB layer turns in-plane. We do not find this behaviour in the simulations and this is therefore still an open question.

As we have shown, the investigated system can be interpreted using a very elementary model that includes all important physical parameters. As similar layer stacks are extremely relevant because they host skyrmions at room temperature~\cite{Luchaire_skyrmion,Woo2016,Hrabec2017,doi:10.1063/1.5048972}, we demonstrate a simple model system to investigate the contributions to the chirality for these systems. We establish the role of dipolar interactions in this article, but this system can also be used for a thorough investigation into the role of the RKKY interaction for the chirality in these systems. Moreover, the thickness dependence of the chirality, demonstrated explicitly in this article, as well as the RKKY interaction that we establish can be very important, suggests two ways in which the chirality in these stacks can be tailored for a specific application. For example, when thinking about magnetic racetrack applications where a uniform chirality of domain walls and skyrmions is preferred we propose two ways to facilitate this.~\cite{Legrandeaat0415,PhysRevB.98.104402,PhysRevApplied.10.064042} We can vary the thickness of the magnetic layers across the stack to alter the competition between the DMI and dipolar interactions on a layer-by-layer basis to modify the position of the transition between the two \Neel{} caps. Second, we can introduce a significant RKKY coupling by modifying the thickness of both Pt and Ir on a sub-nm scale (see Supplementary Note V). By including ferromagnetic RKKY interaction we can make the energy cost for the formation of \Neel{} caps prohibitively high, leading to a uniform chirality determined by the DMI. More generally speaking, the case of anti-ferromagnetic ($J<0$) RKKY interaction is potentially even more interesting, as even without DMI the domain and domain-wall behavior is very rich.~\cite{HELLWIG200713,Lavrijsen2013} Additionally, the method and model system demonstrated here can be extended to further the understanding of the role of chirality in metastable skyrmions towards data storage applications.~\cite{PhysRevApplied.10.064042,PhysRevB.98.104402,Dovzhenko2018}

Concluding, we have shown that by varying the topmost magnetic layer thickness we can tune the relative strength of the DMI with respect to the dipolar coupling which leads to a reversal of  the domain-wall chirality. We believe this to be the first direct demonstration of this competition in the determination of domain-wall chirality. 

This work is part of the research programme of the Foundation for Fundamental Research on Matter (FOM), which is part of the Netherlands Organisation for Scientific Research (NWO). We acknowledge financial support by the DFG within SFB 668. R. A. D. also acknowledges the support of the European Research Council.

J.L. and M.J.M. performed the experiments and data analysis and prepared the manuscript. J.L. performed the analytical computations and micromagnetic simulations. O.K. assisted with the experiments. R.F. and F.K-T. assisted with the startup of the SEMPA system. R.A.D., H.J.M.S., B.K., and R.L. supervised the project. All authors commented on the manuscript.  
\pagebreak
\begin{center}
  \textbf{\large Supplementary material: Tuning magnetic chirality by dipolar interactions}
  \end{center}
  \pagebreak
  
\setcounter{equation}{0}
\setcounter{figure}{0}
\setcounter{table}{0}
\renewcommand{\theequation}{S\arabic{equation}}
\renewcommand{\thefigure}{S\arabic{figure}}
\renewcommand{\thetable}{S\arabic{table}}
\section{Supporting measurements}
In this section we show additional measurements supporting the results of the main paper. First, we show data at additional top CoB layer thicknesses on the same sample as the main paper. We then demonstrate that the measurements are reproducible by showing data on a additional sample that shows the exact same behaviour of domain-wall reversal as a function of top CoB layer thickness.

\begin{figure*}
\centering
\includegraphics{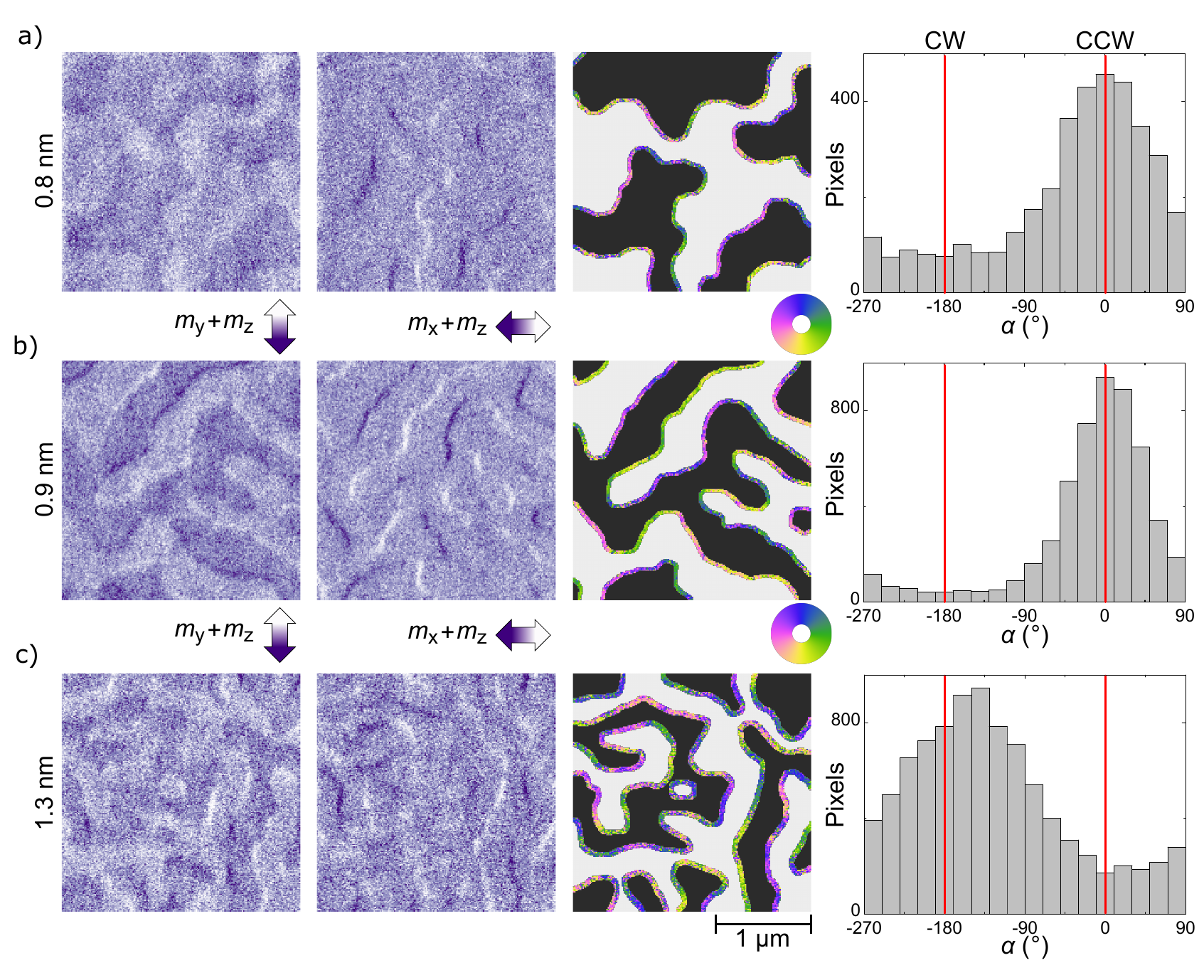}
\caption{\label{fig:sfigure1} SEMPA measurements on the same sample of the main paper, for three different top CoB thicknesses $t$. From a) to c) $t=0.8$, $0.9$ and $1.3$~\si{nm}. See caption of Fig.~1 main paper for more details on what is shown here.}
\end{figure*}

We plot the supporting measurements performed on the same sample as the main paper in~\figref{fig:sfigure1}, where data for thicknesses $t$ of the top CoB layer of $0.8$, $0.9$ and $1.3$~\si{nm} are shown. For the first two thicknesses we find CCW \Neel{} walls. For $t=1.3$~\si{nm} the chirality is reversed and the domain walls are of CW \Neel{} nature. This is in line with the expectations based on the interpretation presented in the main paper.

\begin{figure*}
\centering
\includegraphics{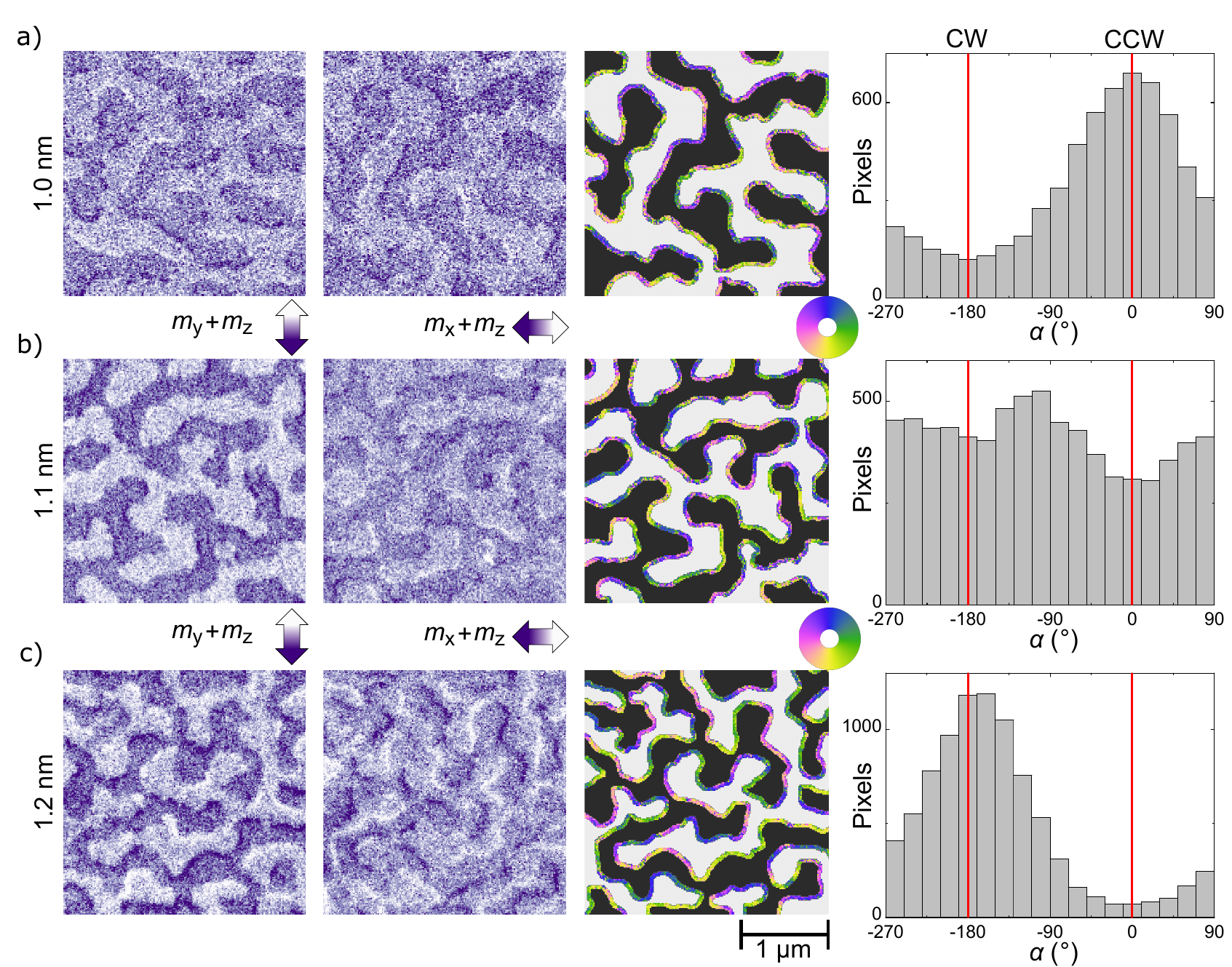}
\caption{\label{fig:sfigure2} SEMPA measurements on a second nominally identical prepared as investigated in the main paper. From a) to c) the thicknesses are $1.0$, $1.1$ and $1.2$~\si{nm}. See caption of Fig.~1 main paper for more details on what is shown here.}
\end{figure*}

In~\figref{fig:sfigure2} we plot an additional sequence of measurements on a nominally identical prepared sample as investigated in the main paper. Here the sample tilt used to get OOP contrast was slightly larger compared to the main paper~\cite{doi:10.1063/1.4998535,PhysRevLett.100.207202} (between $5-10^\circ$ was used compared to $5^\circ$ in the main paper). We find similar behaviour: CCW \Neel{} walls at low thicknesses ($< 1.1$~\si{nm}), a transition point at $1.1$~\si{nm}, and CW \Neel{} walls at larger thicknesses. This demonstrates the reproducibility of the measurements on the same stack.
\section{Details on micromagnetic simulations}
For the micromagnetic simulations, we used~\MuMax{}.~\cite{Vansteenkiste2014} To produce Fig. 3 of the main paper (and~\figref{fig:sfigurerkky_sim}) the following settings were used to run the simulations. The cell sizes are $1\times 8 \times 0.1$~nm (x, y, z) with periodic boundary conditions in the $x$ and $y$ direction equal to 32 repeats. We explicitly introduced spacer layers and the simulation box in the $x$ and $y$ direction was $256\times32$~nm. We used $M_\mathrm{S}=0.49$~\si{MA.m^{-1}} and $K_\mathrm{S}=0.21$~\si{mJ.m^{-2}} given by the result of the SQUID-VSM measurements on a stack without the top CoB layer using the area method to determine the anisotropy.~\cite{0034-4885-59-11-002} We took $A=12$~\si{pJ.m^{-1}}. To our knowledge, there are no exchange constants reported for our CoB composition and layer thickness. We took it slightly smaller than the exchange interaction of $1.6$~\si{pJ.m^{-1}} often assumed for Co~\cite{Eyrich2014,*PhysRevLett.99.217208} because of the reduced correlation number between magnetic atoms in CoB. For each simulation, we initialized two domain walls of square shape with width $5$~\si{nm} in the $x$ direction with orientation $(m_\mathrm{x},m_\mathrm{y},m_\mathrm{z})=(1/\sqrt{2},1/\sqrt{2},0)$ after which we minimized (using default settings) this state to find the equilibrium configuration. For the implementation of the RKKY interaction, we use the method proposed in Ref.~\onlinecite{De_Clercq_2017}, where we add custom magnetic fields across non-magnetic spacer layers to act as RKKY fields.

We have verified that the exact parameters used for the simulations did not significantly affect the outcome of the simulations. We changed the x-cell size to $0.25$~\si{nm} and saw no change in the phase diagram. Similarly, by changing the amount of periodic repeats to $64$ in each direction, the phase diagram was unaltered. We further checked the stopping conditions for the minimizer (from $\delta\boldsymbol{m}=10^{-6}$ to $10^{-7}$, numerical noise prevented us from going down any further) and the changes were minimal. Increasing the simulation box in the $x$-direction to $512$~\si{nm} did show a small effect. Mostly for thicker top CoB layers $t=1.3,1.4$~\si{nm} where a shift in transition DMI of about $-0.05$~\si{pJ.m^{-1}} can be found, as well as a slightly wider transition region at $1.4$~\si{nm}. However, this also shows up in the analytical model through its dependence of $\lambda$ (see note III) and is the result of the stray fields being domain size dependent. 
\section{Model calculations}
In this section we adapt the model introduced in Ref.~\onlinecite{Legrandeaat0415} to our situation. For an overview and derivation of the equations presented here, we refer the reader to the original work. We will only present the end results for the modified situation here. Additionally, we will present phase diagrams to demonstrate the effect of a varying domain-wall width of the top layer as a function of $t$.

Compared to the original model we now assume that the domain wall in the top layer has a width $\varDelta_\mathrm{top}$ and the domain walls in the bottom layers can have a different width $\varDelta_\mathrm{bot}$ that is equal across all these layers. We further change the constraint that all layers have equal thicknesses. The top layer now has thickness $t$ and the bottom layers all have equal thickness $q$. We furthermore add an additional field from the RKKY interaction that is not present in the original work.

Following a trivial alteration of the derivation we now modify the field from the surface charges of the underlying layers to
\begin{equation}
B_\mathrm{dip,S}=-\mu_\mathrm{0}\sum_{k=1,\mathrm{odd}}^{\infty} \frac{4\pi M_\mathrm{S}\varDelta_\mathrm{bot}}{\lambda}\mathrm{csch}\bigg( \frac{\pi^2 \varDelta_\mathrm{bot} k}{\lambda}\bigg)\mathrm{sinh}\bigg(\frac{\pi k q}{\lambda}\bigg)  \mathrm{cos}\bigg(\frac{2 \pi k x}{\lambda}\bigg) \mathrm{e}^{-\frac{\pi k (t-q)}{ \lambda}}  \frac{\mathrm{e}^{-\frac{2\pi k p}{\lambda}}-\mathrm{e}^{-\frac{2\pi k p N}{\lambda}}}{1-\mathrm{e}^{-\frac{2\pi k p}{\lambda}}},
\end{equation}
with $N$ the total amount of magnetic layers in the system (including the top one), $p$ the periodicity of the underlying stack ($q$ + thickness of spacer layer) and  $\lambda$ the domain periodicity. A similar analysis for the field from the volume charges of the underlying layer gives
\begin{equation}
B_\mathrm{dip,V}=-\mu_\mathrm{0}\sum_{k=1,\mathrm{odd}}^{\infty} \frac{4\pi M_\mathrm{S}\varDelta_\mathrm{bot}}{\lambda}\mathrm{sech}\bigg( \frac{\pi^2 \varDelta_\mathrm{bot} k}{\lambda}\bigg)\mathrm{sinh}\bigg(\frac{\pi k q}{\lambda}\bigg)  \mathrm{cos}\bigg(\frac{2 \pi k x}{\lambda}\bigg) \mathrm{e}^{-\frac{\pi k (t-q)}{ \lambda}}  \frac{\mathrm{e}^{-\frac{2\pi k p}{\lambda}}-\mathrm{e}^{-\frac{2\pi k p N}{\lambda}}}{1-\mathrm{e}^{-\frac{2\pi k p}{\lambda}}},
\end{equation}
and we can modify the demagnetizing field of the top layer itself to
\begin{equation}
B_\mathrm{dip,self}=\mu_\mathrm{0}\sum_{k=1,\mathrm{odd}}^{\infty} \frac{4\pi M_\mathrm{S}\varDelta_\mathrm{top}}{\lambda}\mathrm{sech}\bigg( \frac{\pi^2 \varDelta_\mathrm{top} k}{\lambda}\bigg)  \mathrm{cos}\bigg(\frac{2 \pi k x}{\lambda}\bigg) \Bigg[\frac{1-\mathrm{e}^{-\frac{2\pi k  t}{\lambda}}}{2} \frac{\lambda}{\pi k t}-1 \Bigg].
\end{equation}
Lastly, we add to this model an additional RKKY field as
\begin{equation}
B_\mathrm{RKKY}=\frac{J}{2 M_\mathrm{S} t} \mathrm{sech}\bigg(\frac{x}{\varDelta_\mathrm{bot}} \bigg),
\end{equation}
with $J$ the RKKY coupling strength. Of course, in the determination of the DMI field (see Ref.~\onlinecite{Legrandeaat0415} for details on this field) we use the thickness dependence as determined by $t$ and when integrating these fields across the domain-wall profile we use the profile of the top layer with width $\varDelta_\mathrm{top}$.

For the phase diagrams presented in the main paper, we match the domain-wall width of the top layer $\varDelta_\mathrm{top}$ to the top layer of the micromagnetic simulation by fitting the in-plane component of the domain wall with a $\mathrm{sech}(x/\varDelta)$ function. The micromagnetic parameters used for the simulations were $t=1.2$~\si{nm}, $D=0.6$~\si{pJ.m^{-1}} and $J=0$~\si{mJ.m^{-2}}. We matched the domain-wall width of the underlying layers $\varDelta_\mathrm{bot}$ to the domain-wall width of that same simulation for the bottom layer. This gives $\varDelta_{\mathrm{top}}=14.4$~\si{nm} and $\varDelta_{\mathrm{bottom}}=9.1$~\si{nm}.  Furthermore, we use $\lambda=256$~\si{nm} to match the periodicity of the micromagnetic simulations. Lastly, we used $M_\mathrm{S}=0.49$~\si{MA.m^{-1}} determined by SQUID-VSM measurements on a stack without the top CoB layer.

\begin{figure*}
\centering
\includegraphics{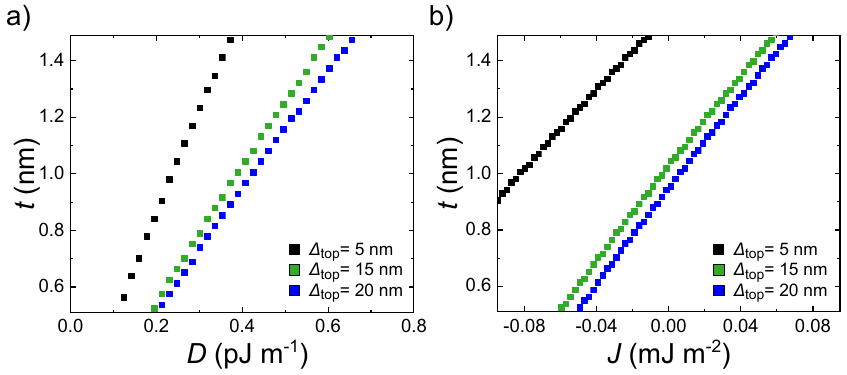}
\caption{\label{fig:deltacomp} a) Transition thickness as a function of $D$ (a) and $J$ (b) for different top domain-wall widths $\varDelta_\mathrm{top}$. The rest of the parameters are the same as Fig.~2c and d of the main paper. }
\end{figure*}

In Fig.~2c and d of the main paper we present the main phase diagrams that result from this model, where we have a transition from CW to CCW \Neel{} walls as a function of $D$, $J$ and $t$, assuming the top domain-wall width is constant throughout the calculations. However, we know from sec.~VII of this supplementary material that this assumption is invalid. So, for clarity we show in~\figref{fig:deltacomp}a and b what happens to these phase diagrams when we vary this top domain-wall width. We see that the transition thickness goes up both as a function of $D$ and $J$ for smaller domain-wall widths. As the integrated DMI field is independent of the domain-wall width, this is the result of a reduced integrated effective field of the RKKY and dipolar fields due to this smaller domain-wall width. This then explains the slight mismatch between the micromagnetic simulations and the model as shown in Fig.~3c of the main paper and~\figref{fig:sfigurerkky_sim} of this supplementary material, as the dependence of $\varDelta_{\mathrm{top}}$ on $t$ is not taken into account.

\section{Histogram width}
In this section we demonstrate that the majority of the width in the histograms is given by errors in the determination of the domain-wall normal $n$ and the magnetization direction $m$, rather than an actual spread in domain-wall angles $\alpha$.
\begin{figure*}
\centering
\includegraphics{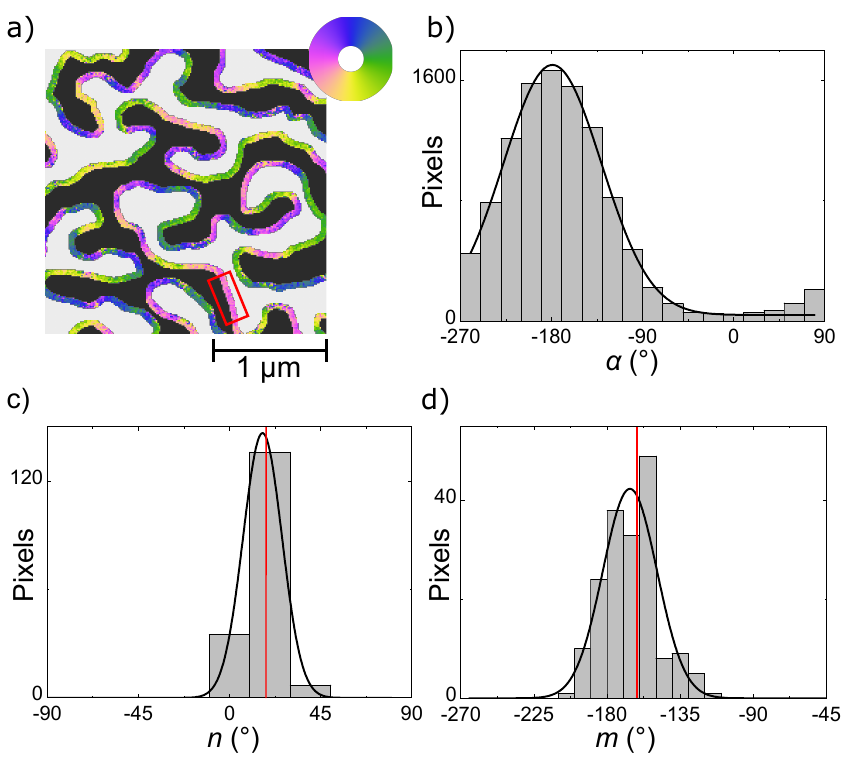}
\caption{\label{fig:histobroad} a) Composite SEMPA image for $t=1.2$~nm (see main paper) with the domain wall analysed indicated with the red box. b-d) Histogram of the domain-wall angle $\alpha$ (b), angle of the domain-wall normal $n$ (c), and angle of magnetization $m$ (d) including fits with a Gaussian. The expected value is indicated in red (for c dictated by approximate wall normal and for d assuming a \Neel{} wall). For b) the histogram is taken for all the pixels in all walls in a) and for c,d) the histogram is taken only for the wall indicated in red in a).}
\end{figure*}

To give an estimate of both contributions to the width of the histogram, we look at a relatively straight piece of domain wall, such as for example the wall indicated by the red box in~\figref{fig:histobroad}a for $t=1.2$~\si{nm}. For this relatively straight piece of wall we can determine the spread in domain-wall normal $n$ (\figref{fig:histobroad}c) and magnetization $m$ (\figref{fig:histobroad}d). We fitted these histograms and extracted values for the full width half maximum (FWHM) which we give in Table.~\ref{Tab:table1}. Also included in this table is a value for the FWHM of the complete histogram for the angle $\alpha$ for all pixels in the walls (see Fig.~1 of the main paper), which was also obtained using a fit (\figref{fig:histobroad}b). In the table we further indicate values extracted in a similar way for $2$ other thicknesses.

For this short and relatively straight piece of domain wall the distribution of the domain-wall normal $n$ and magnetization $m$ should resemble a delta peak. However, we find a broad histogram for both $n$ and $m$ in \figref{fig:histobroad} for this short domain wall because of errors in determining the domain-wall normal and Poisson noise due electron counting~\cite{doi:10.1063/1.3534832} in the determination of the magnetization direction $m$. If we compare the FWHM for this distribution of $n$ and $m$ of to the FWHM of $\alpha$, as we do in Table.~\ref{Tab:table1}, we then conclude the majority of the FWHM in $\alpha$ is dictated by these errors. This means that the width of the histogram is no indication of a spread in actual domain-wall angles in our samples. 

\begin{table}[ht]
\centering
\caption{\label{Tab:table1} Values of the FWHM of the domain-wall angle $\alpha$ , angle of the domain-wall normal $n$ and angle of magnetization $m$ for different top CoB thicknesses $t$.}
\setlength{\tabcolsep}{0.75em}
\begin{tabular}[t]{lccc}
\hline
$t$~(\si{nm})   &    $n$~($^\circ$)   		&$m$~($^\circ$)		&  $\alpha$~($^\circ$) \\
\hline
$0.9$   &    $21.0 \pm 0.9$		& $52 \pm 2$ 	& $89 \pm 2$\\
$1.0$&    $43 \pm 2$			& $63 \pm 5$ 	& $111 \pm 4$\\
$1.2$   &    $22.65 \pm 0.02$	& $40 \pm 3$ 	& $114 \pm 3$\\
\hline
\end{tabular}
\end{table}%

\section{Experimental RKKY coupling}
In this section we prove that the RKKY coupling in our measured sample is small and can be safely neglected when interpreting the data. However, we also show that by slightly tuning the Pt and Ir layer thickness we can significantly enhance the coupling to a point that we need to take it into account. See the discussion in the main paper for further details.
\begin{figure*}
\centering
\includegraphics[width=\columnwidth]{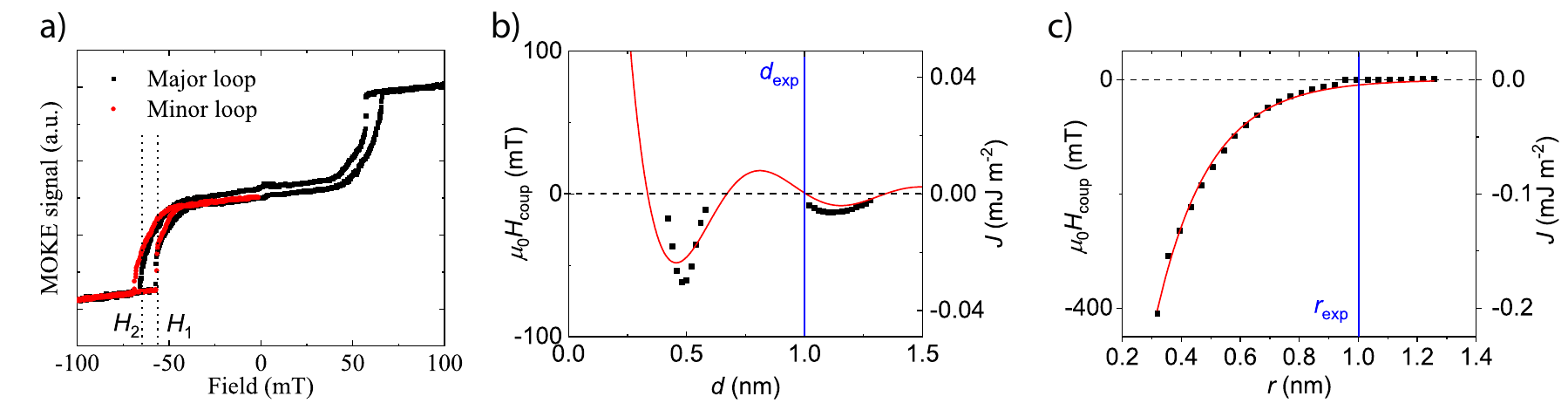}
\caption{\label{fig:sfigurerkky_exp} a) Major and minor hysteresis loop for $d=0.5$~\si{nm} and $r=0.5$~\si{nm} measured using MOKE. Also indicated are the switching fields $H_\mathrm{1}$ and $H_\mathrm{2}$ used to determine the RKKY interaction. b) Coupling field $H_\mathrm{coup}$ (left axis) and RKKY strength $J$ (right axis) as a function of Ir thickness $d$ for $r=0.5$~\si{nm} including a fit. c) Coupling field $H_\mathrm{coup}$ (left axis) and RKKY strength $J$ (right axis) as a function of Pt thickness $r$ at $d=0.48$~\si{nm} including a fit with an exponential decay. For b) and c) the blue lines indicate the thicknesses of the corresponding layers for the investigated samples in the main paper.}
\end{figure*}

We measured the RKKY coupling in Ta($4$)/Pt($3$)/Co$_{\mathrm{80}}$B$_{\mathrm{20}}$/Ir($d$)/Pt($r$)/Co$_{\mathrm{80}}$B$_{\mathrm{20}}$/Ir($1$)/Pt($4$) stacks using MOKE and by looking at the switching fields.~\cite{doi:10.1063/1.3682103} We first investigate the RKKY coupling as a function of Ir thickness $d$ on a wedged sample with $r=0.5$~\si{nm} because we find that the coupling is negligible at a Pt thickness $r$ of $1.0$~\si{nm}. We will demonstrate the effect of an increased Pt thickness $r$ later in this section. A MOKE hysteresis loop on such a stack with $d=0.5$~\si{nm} is shown in~\figref{fig:sfigurerkky_exp}a. The double switch in the major loop is indicative of an antiferromagnetic coupling between the two CoB layers. To determine the RKKY interaction, we take a minor loop measuring the first switch only (also shown) and determine the two switching fields $H_\mathrm{1}$ and $H_\mathrm{2}$. The coupling field $H_\mathrm{coup}$ is then calculated as $(H_\mathrm{1}+H_\mathrm{2})/2$ and the RKKY interaction $J$ as $\mu_\mathrm{0} H_\mathrm{coup} M_\mathrm{S}t_\mathrm{CoB}$. Doing this as a function of Ir thickness gives us~\figref{fig:sfigurerkky_exp}b where we include a fit with the predicted $A\,\mathrm{sin}(k d)/d^2$ dependence.~\cite{PhysRevB.52.411} Although there are no hard switches in our minor loops that are typically used to extract $J$, we believe that our analysis yields representative results as the first AF peak lies at $\sim 0.5$~\si{nm}, in full agreement with literature.~\cite{HELLWIG200713}

At the experimental thickness of Ir ($d_\mathrm{exp}$) we are at a root of the RKKY coupling, suggesting the RKKY coupling is negligible in the sample investigated in the main paper. Furthermore, the overall coupling in~\figref{fig:sfigurerkky_exp}b  is of the order of $0.02$~\si{mJ.m^{-2}}. However, this is at a Pt thickness of $0.5$~\si{nm}. We expect this coupling to go down as a function of Pt thickness,~\cite{doi:10.1063/1.3682103} and this is indeed what we observe in~\figref{fig:sfigurerkky_exp}c, where we plot the coupling as function of the Pt thickness on sample where we wedged the Pt thickness $r$ and kept Ir at $d=0.5$~\si{nm}. The coupling indeed decreases exponentially with a decay length around $0.2$~\si{nm} and is negligible at a Pt thickness of $1.0$~\si{nm}. As this is the thickness for the samples investigated by SEMPA, this gives us another reason to neglect the RKKY coupling in our interpretation. 

However, we also see here that a ferromagnetic RKKY coupling with a value of $0.01$~\si{mJ.m^{-2}} would require a Pt thickness of around $0.5$~\si{nm} and an Ir thickness of $0.75$~nm. With these parameters, the RKKY coupling should have a significant contribution to the domain-wall chirality and shift the transition thickness $t_\mathrm{tr}$ to higher values.

\section{Simulations RKKY coupling}
In Fig.~3c of the main paper we compare the micromagnetic simulations to the analytical model as a function of $D$ and $t$. We can also compare the analytical calculations as a function of the RKKY interaction, shown in in Fig.~2d of the main paper, to the micromagnetic simulations
\begin{figure*}
\centering
\includegraphics{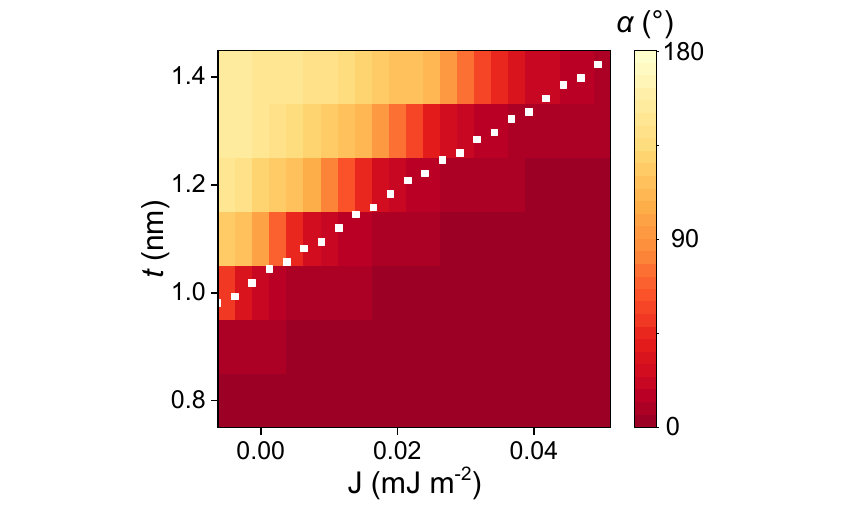}
\caption{\label{fig:sfigurerkky_sim} Phase diagram for the domain-wall angle $\alpha$~(see Fig.~1c of main paper) in the top CoB layer from micromagnetics as a function of $J$ and $t$ with $D=0.4$~\si{pJ.m^{-1}}. Also indicated is the transition line from CW to CCW from Fig.~2d of the main paper.}
\end{figure*}

This comparison is shown in~\figref{fig:sfigurerkky_sim}, where we plot the domain-wall angle $\alpha$ of the domain wall in the top CoB layer from the micromagnetic simulations as a function of $J$ and $t$, together with the transition line from CW to CCW \Neel{} walls indicated with the white dots. The domain-wall chirality reverses between CCW ($\alpha=0^\circ$) and CW ($\alpha=180^\circ$) as a function of $J$ and $t$, which matches the analytical calculations. Additionally, similar to the difference between simulations and calculations as a function of $D$, also here we find a region with Bloch domain walls ($\alpha=90^\circ$). The slight discrepancy between simulations and calculations can be explained by the variation in the domain-wall width as a function of top CoB thickness, as we demonstrated in Sec.~III of this supplementary.
 
\section{Domain-wall width}
In this section we extract the domain-wall width from the SEMPA measurements and compare them to the domain-wall widths determined from the micromagnetic simulations. We demonstrate that the domain-wall widths are in agreement with each other, and that the simulations demonstrate that the domain-wall width does indeed vary as a function of top CoB thickness $t$.

\begin{figure*}
\centering
\includegraphics{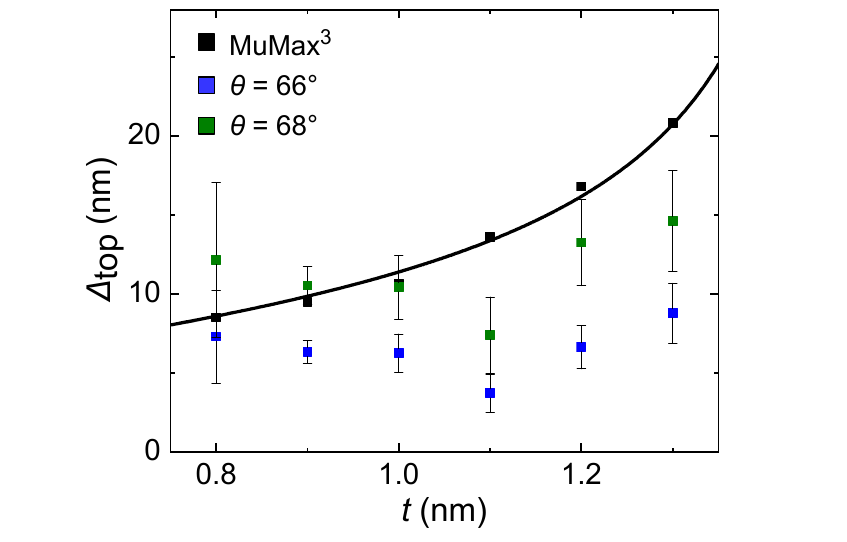}
\caption{\label{fig:swidth} Domain-wall width $\varDelta_\mathrm{top}$ as a function of top CoB layer thickness $t$ for micromagnetic simulations and the experimental situation. We plot the experimental situation for two different assumed sample tilt angles $\theta$. The line is a guide to the eye for the \MuMax{} simulations.}
\end{figure*}

Because the electron beam spot size is much larger than the domain-wall width, we need to be careful when estimating the domain-wall width $\varDelta_\mathrm{top}$. Following Ref.~\onlinecite{PhysRevB.96.060410}, we fit the domain-wall profile with a Gaussian distribution. This then allows us to estimate $\varDelta_\mathrm{top}$ when we have a value for the intrinsic SEMPA asymmetry $A$. We estimate this $A$ by determining the OOP contrast in the $m_\mathrm{y}$ scan (average of all domains). This is then converted to a situation of pure IP contrast by dividing it by $\mathrm{sin}(\phi)$ with $\phi$ the tilt angle with respect to the SEMPA detector (which is a few degrees). For the micromagnetic domain walls, we extract $\varDelta_\mathrm{top}$ by fitting the average OOP magnetzation to the following function: $\mathrm{tanh}(x/\varDelta_\mathrm{top})$.

The resulting domain-wall widths $\varDelta_\mathrm{top}$ are plotted in~\figref{fig:swidth}, where we also plot the experimental $\varDelta_\mathrm{top}$ for two different sample tilt angles $\theta$ (more on this in the next paragraph). The micromagnetic simulations provide domain-wall widths that increase with the top CoB thickness. However, with the error bars present on the experimental data we cannot conclude from this alone that the domain-wall width increases with $t$ for all thicknesses investigated. However, we find that the average domain-wall width of the SEMPA measurements and \MuMax{} simulations are in excellent agreement.

The tilt angle $\theta$ should correspond to a tilt of our sample-stage with respect to the SEM column when the OOP contrast should be zero (i.e. $\phi=0$). As this angle depends very much on the exact sample mounting conditions (we estimate $\theta=65^\circ$), it is difficult to know this exactly. However, as it affects $\phi$ we need to take this into account because $\phi$ has a significant effect on the domain-wall width $\varDelta$. From \figref{fig:swidth} we conclude that the experimental data at $\theta=68^\circ$ matches the micromagnetic simulations well.

\pagebreak

\bibliography{references}
\end{document}